\documentclass[twocolumn]{svjour3}          
\smartqed  
\usepackage{graphicx}
\usepackage{amsmath}
\usepackage{cite}
\usepackage[pdfborder={0 0 0},colorlinks=true,linkcolor=blue,citecolor=blue]{hyperref}

\def\beq{\begin{equation}}
\def\eeq{\end{equation}}
\def\bea{\begin{eqnarray}}
\def\eea{\end{eqnarray}}

\def\q{\mathbf{q}}
\def\k{\mathbf{k}}
\def\p{\mathbf{p}}
\def\rr{\mathbf{r}}

\def\K{\mathbf{K}}

\newcommand{\su}{\uparrow}
\newcommand{\sd}{\downarrow}

\newcommand{\ii}{{\mathrm{i}}}

\newcommand{\nn}{\nonumber}

\begin{document}

\title{Spin and charge susceptibilities of the two-orbital model within the cluster perturbation theory for Fe-based materials}

\titlerunning{Spin and charge susceptibilities of the two-orbital model within the CPT for Fe-based materials}  

\author{S.V. Nikolaev \and M.M. Korshunov}


\institute{S.V. Nikolaev and M.M. Korshunov \at
              Kirensky Institute of Physics, Akademgorodok, Krasnoyarsk 660036, Russia\\
              \email{mkor@iph.krasn.ru}
              \and
              Siberian Federal University, Svobodny Prospect 79, Krasnoyarsk 660041, Russia
}

\date{Received: date / Accepted: date}

\maketitle

\begin{abstract}
Cluster perturbation theory is used to calculate band structure, spectral functions, Fermi surface, and spin and charge susceptibilities for the two-orbital model of iron pnictides with the on-site multiorbital Hubbard interactions. Susceptibilities are calculated within the approximation combining the cluster perturbation theory for the self-energy corrections and the random-phase approximation (RPA) for the vertex renormalizations. Calculations for the small values of Hubbard repulsion $U \leq 2$~eV confirm that the rigid band approximation and RPA for the spin and charge susceptibilities are suitable approaches for the case of weak interactions.
\end{abstract}

\section{Introduction}

Iron-based materials - pnictides and chalcogenides - represent a new class of unconventional superconductors with high transition temperatures~\cite{Kamihara2008,Sadovskii2008,Paglione2010,Johnston2010,Mazin2010,Wen2011,Stewart2011,ROPPreview2011,Hirschfeld2016}. While the mechanism of superconductivity is still a mystery, the main candidates are spin and orbital fluctuations. Except for the extreme hole and electron dopings, the Fermi surface (FS) consists of two or three hole pockets around the $\Gamma=(0,0)$ point and two electron pockets around the $X=(\pi,0)$ and $Y=(0,\pi)$ point in the Brillouin zone, corresponding to the one iron per unit cell. Different mechanisms of Cooper pairs formation result in the distinct superconducting gap symmetry and structure in Fe-based superconductors (FeBS)~\cite{ROPPreview2011}. For example, spin fluctuation approach leads to the extended $s_\pm$ state ($s$-wave gap that changes sign between hole and electron FSs) as the main instability~\cite{Mazin_etal_splusminus,Graser2009,Kuroki2008,Maiti,MaitiPRB,KorshunovUFN}, while orbital fluctuations promote the order parameter to have the sign-preserving $s_{++}$ symmetry~\cite{Kontani2010,Onari2012}.

Most approaches to the superconductivity theory in FeBS including spin fluctuations in the random-phase approximation (RPA) are solidly justified in the case of a weak interaction between electrons. Agreement between the experimental FS and the one theoretically obtained within density functional theory (DFT) as well as the smallness of the magnetic moment in most FeBS and absence of Mott insulating state even in the undoped materials assert that the interaction is weak. On the other hand, comparison of ARPES (angle-resolved photoemission spectroscopy) results and DFT bands shows the bandwidth reduction about two to three times~\cite{Kordyuk2012}, and the redistribution of spectral wight from the Drude peak to higher energies in optical conductivity of LaFePO and BaFe$_2$As$_2$ points out to the at least moderate electronic correlations~\cite{Qazilbash2009}. While the use of hybrid methods like LDA+DMFT (local density approximation + dynamical mean-field theory) to treat electronic correlations allows to describe some physical properties of FeBS~\cite{Haule2008,Skornyakov2009,Skornyakov2010,Yin2011}, nonlocal spin fluctuations are beyond these approaches. Thus it is hard to justify use of methods like LDA+DMFT to build up a theory of superconductivity where spin fluctuations are crucial~\cite{ROPPreview2011}. Cluster extensions of DMFT, e.g., CDMFT (cellular DMFT)~\cite{Kotliar2001} and DCA (dynamical cluster approximation)~\cite{Hettler1998,Lichtenstein2000,Maier2005}, are numerically very expensive for the multiorbital systems. Here we use alternative approach called the cluster perturbation theory (CPT)~\cite{Senechal2000,Senechal2002}. It relays on the exact diagonalization of the small cluster to calculate a cluster Green's function. Then the intercluster hoppings and interactions are treated as perturbations. Such procedure allows to describe spin and charge fluctuations within the cluster exactly.

Here we use CPT to calculate spin and charge susceptibilities for the simple two-orbital model of iron pnictides~\cite{Raghu2008} with the on-site multiorbital Hubbard interactions. First, we calculate full Green's functions via CPT. Then susceptibilities are obtained within RPA with the bare polarization bubble composed of full CPT Green's functions. Since the susceptibility is the central part of the spin/charge fluctuation-driven Cooper pairing, this is the essential step towards the theory of superconductivity in FeBS.

\section{Model}

\begin{figure}[t]
\begin{center}
\includegraphics[width=.47\textwidth]{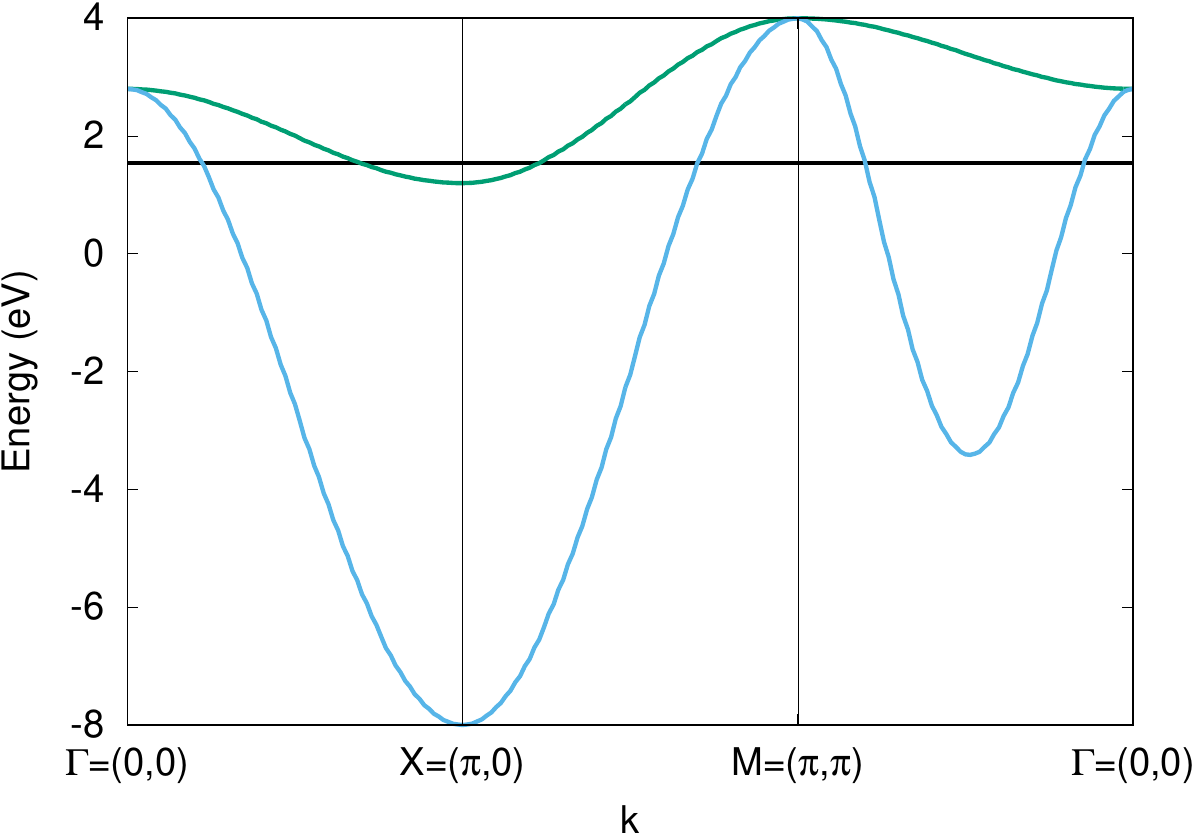}
\caption{Band structure of model~(\ref{eq:H0}). Position of the chemical potential is marked by the black horizontal line.
\label{fig:ek}}
\end{center}
\end{figure}

\begin{figure}[t]
\begin{center}
\includegraphics[width=.3\textwidth]{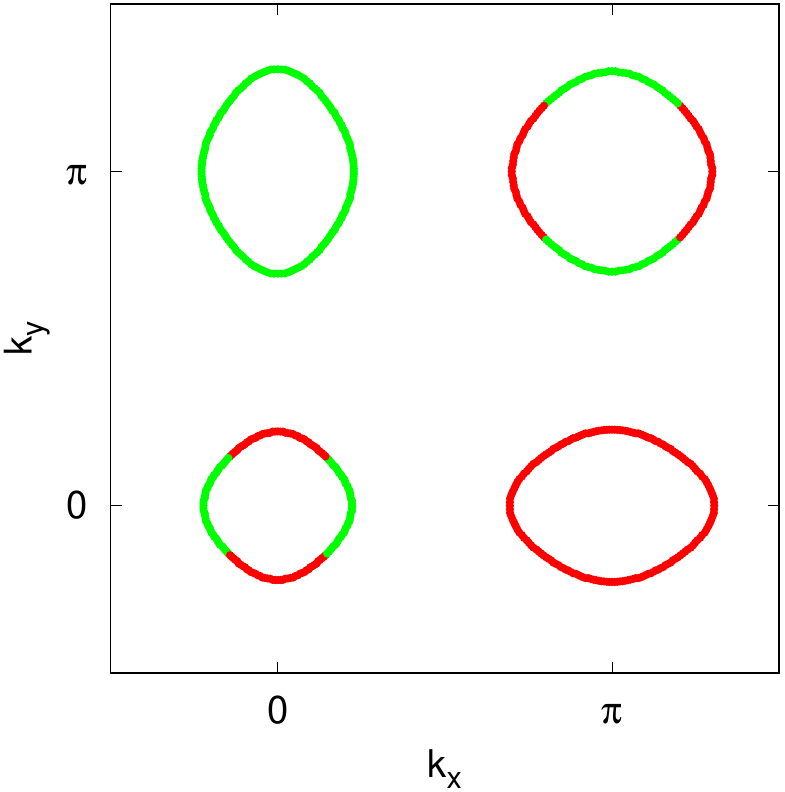}
\caption{Fermi surface of model~(\ref{eq:H0}). Different colors indicate major orbital contribution (red - $d_{xz}$, green - $d_{yz}$).
\label{fig:fs}}
\end{center}
\end{figure}

To preserve orbital content of the bands and still gain some simplicity, we study here the two-orbital tight-binding model from Ref.~\cite{Raghu2008} with Hamiltonian
\begin{equation}
 H_0 = \sum_{\k \sigma} \sum_{l l'} \left[ t_{l l'}(\k) + \epsilon_{l} \delta_{l l'} \right] d_{l \k \sigma}^\dagger d_{l' \k \sigma},
 \label{eq:H0}
\end{equation}
where $d_{l \k \sigma}^\dagger$ is the annihilation operator of a particle with momentum $\k$, spin $\sigma$, and orbital index $l = 1,2$ enumerating $d_{xz}$ and $d_{yz}$ orbitals. Later we use numerical values of hopping matrix elements $t_{l l'}(\k)$ and one-electron energies $\epsilon_{l}$ from Ref.~\cite{Raghu2008}. This model produce the band structure shown in Fig.~\ref{fig:ek} and the FS composed of one hole pocket around the $\Gamma$ point and two electron pockets centered around $X$ and $Y$ points, see Fig.~\ref{fig:fs}.
%
%
The model can be used to describe the two electronic components scenario~\cite{Bianconi2013} where the bottom or the top of one of the bands is close to the chemical potential~\cite{Kordyuk2012}. Similar scenario with the proximity to a Lifshitz transition has been proposed earlier for cuprates~\cite{Bianconi2000}.


The general two-particle on-site interaction is given by the Hamiltonian~\cite{Castallani1978,Oles1983,Kuroki2008,Graser2009}:
\bea
H_{int} &=& U \sum_{f, m} n_{f m \su} n_{f m \sd} + U' \sum_{f, m < l} n_{f l} n_{f m} \nn\\
  && + J \sum_{f, m < l} \sum_{\sigma,\sigma'} d_{f l \sigma}^\dag d_{f m \sigma'}^\dag d_{f l \sigma'} d_{f m \sigma} \nn\\
  && + J' \sum_{f, m \neq l} d_{f l \su}^\dag d_{f l \sd}^\dag d_{f m \sd} d_{f m \su}.
\label{eq:Hint}
\eea
where $n_{f m} = n_{f m \su} + n_{f m \sd}$, $n_{f m \sigma} = d_{f m \sigma}^\dag d_{f m \sigma}$ is the number of particles operator at the site $f$, $U$ and $U'$ are the intra- and interorbital Hubbard repulsion, $J$ is the Hund's exchange, and $J'$ is the so-called pair hopping. We restrict the number of interaction parameters by obeying the spin-rotational invariance: $U'=U-2J$ and $J'=J$.

In principle, the phase separation can appear in the two-band Hubbard model as has been shown in Ref.~\cite{Kugel2008}. However, here we do not consider it.

\section{Cluster perturbation theory}

To study the interacting system with the Hamiltonian $H = H_0 + H_{int}$, we use the cluster perturbation theory~\cite{Senechal2000,Senechal2002}. First step is the exact diagonalization of the small cluster. Here we choose $2 \times 2$ cluster of iron sites. Each site has 2 iron orbitals. The initial lattice is tiled by identical $2 \times 2$ clusters. Thus, the lattice transforms into a superlattice of clusters with a new translational order of an artificial origin. To avoid artificial splitting of energy bands, here we use the averaging over two different tiling patterns that is discussed in the framework of the norm-conserving cluster perturbation theory (NC-CPT)~\cite{Nikolaev2010,Ovchinnikov2011,Nikolaev2012}. We treat the intercluster hoppings and interactions as perturbations.

As the first step, we calculate the cluster Green's function $G_{il,jm}^{(c)}(\omega)$ by the Lanczos algorithm. Here $i$ and $j$ are intracluster site indices, $l$ and $m$ are orbital indices. The next step is to determine the full matrix Green's function,
\begin{equation}
 \hat{G}^{-1}(\K, \omega) = \hat{G}^{(c)-1}(\omega) - \hat{V}(\K),
 \label{eq:full_G}
\end{equation}
where $\K$ is a wave vector in the reduced Brillouin zone (the Brillouin zone of the superlattice), and $\hat{V}(\K)$ is the matrix of the intercluster hoppings defined as
\begin{equation}
 V_{il,jm}(\K) = \sum_{h} t^{h}_{il,jm} e^{\ii \K \rr_{h}}.
 \label{eq:mat_V}
\end{equation}
Here $\rr_{h} = \rr_{g} - \rr_{g'}$ with $\rr_{g}$ being the radius vector of the neighboring clusters labelled $g$.

To restore the full translation symmetry of the lattice we perform a residual Fourier transform and obtain a momentum dependent Green's function in orbital basis,
\begin{equation}
 G_{l m}(\k,\omega) = \frac{1}{N_c} \sum_{i,j=1}^{N_c} G_{il,jm}(\k,\omega) e^{-\ii \k (\rr_i - \rr_j)},
 \label{eq:fun_G}
\end{equation}
where $\rr_i$ is the radius vector of the site $i$ within the cluster, $N_c$ is the number of sites in the cluster (which is four in our case), and $\k$ is a wave vector in the Brillouin zone of the initial lattice. Here we used a translational invariance of the intercluster hoppings matrix, $\hat{V}(\K) = \hat{V}(\k)$.


\section{Susceptibility calculation}

Transverse component of the bare spin susceptibility that is a tensor in orbital indices $l$, $l'$, $m$, $m'$ have the following form~\cite{KorshunovUFN},
\begin{eqnarray}\label{eq:chipm}
 &&\chi_{ll',mm'}^{(0)}(\q,\ii\Omega) \nn\\
 &&= -T \sum_{\p,\omega_n} G_{m l \su}^{(0)}(\p,\ii\omega_n) G_{l' m' \sd}^{(0)}(\p+\q,\ii\Omega+\ii\omega_n).
\end{eqnarray}
Here $\Omega$ and $\omega_n$ are Matsubara frequencies, and $G_{l m \su}^{(0)}(\p,\ii\omega_n)$ is the Green's function of the noninteracting system~(\ref{eq:H0}).

\begin{figure}[t]
\begin{center}
\includegraphics[width=.48\textwidth]{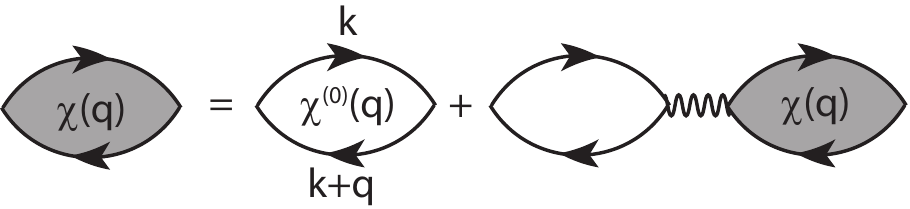}
\includegraphics[width=.28\textwidth]{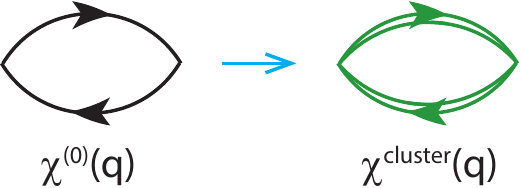}
\caption{Diagrammatic representation of the RPA equation for the susceptibility $\chi(\q,\omega)$ (top) and the renormalization of the Green's function forming the bare polarization bubble $\chi^{(0)}(\q,\omega)$ by the CPT Green's functions (bottom). Wavy line represents two-body interaction~(\ref{eq:Hint}).
\label{fig:chidiagrams}}
\end{center}
\end{figure}

We can now make a replacement
$G_{l m \su}^{(0)}(\p,\ii\omega_n) \to G_{l m \su}(\p,\ii\omega_n)$,
and instead of $\chi_{ll',mm'}^{(0)}$ we will have $\chi_{ll',mm'}^{\mathrm{cluster}}$ with the Green's functions obtained via CPT. Thus we retain intracluster self-energy corrections but loose the long tail of intercluster effective interaction. To overcome this drawback we use RPA series with the ``cluster'' susceptibility $\chi_{ll',mm'}^{\mathrm{cluster}}$ replacing the bare electron-hole bubble, see Fig.~\ref{fig:chidiagrams}. There is no double-counting problem here since the cluster susceptibility includes only self-energy corrections and RPA is the vertex renormalization.

Therefore, RPA susceptibility $\chi_{ll,mm}(\q,\ii\Omega)$ is obtained by solving the equation that is shown graphically in Fig.~\ref{fig:chidiagrams} with the interaction represented by matrix $U_s$ for spin and $U_c$ for charge susceptibility. Exact form of these matrices for Hamiltonian~(\ref{eq:Hint}) is given in Ref.~\cite{Graser2009}. To use matrix notations we introduce the correspondence between matrix ($\imath$, $\jmath$) and orbital indices: $\imath = l + l' n_O$ and $\jmath = m + m' n_O$, where $n_O=2$ is the number of orbitals.

Here we use the continuation of cluster Green's functions to Matsubara frequencies via spectral representation,
\begin{equation}
 G_{l m}(\k,\ii\omega_n) = \int\limits_{-\infty}^{+\infty} d\omega' \frac{A_{l m}(\k,\omega')}{\ii\omega_n - \omega'},
 \label{eq:mat_fun_G}
\end{equation}
where $A_{l m}(\k,\omega) = -\frac{1}{\pi} \mathrm{Im} G_{l m}(\k,\omega)$ is the spectral function.

After substitution of (\ref{eq:mat_fun_G}) into (\ref{eq:chipm}) and summation over $\omega_n$ we obtain the cluster susceptibility,
\begin{eqnarray}
 \chi_{ll',mm'}^{\mathrm{cluster}}(\q,\ii\Omega) = &&-\sum_{\p} \iint\limits_{-\infty}^{+\infty} d\omega' d\omega'' A_{ml}(\p,\omega') \times \nn\\
 &&\times A_{l'm'}(\p + \q,\omega'') \frac{f(\omega') - f(\omega'')}{\omega' - \omega'' + \ii\Omega},
 \label{eq:chipm_cluster}
\end{eqnarray}
where $ f(\omega) = 1/\left[1 + e^{(\omega - \mu)/T}\right]$ is the Fermi function. After the calculation, we make the analytical continuation to real frequencies, $\ii\omega_n \to \omega + \ii\delta$ with $\delta \to +0$.

Physical spin susceptibility is given by the trace over orbital indices, $\chi(\q,\omega) = \frac{1}{2} \sum_{l,m} \chi_{ll,mm}(\q,\omega)$.

\section{Numerical results}

\begin{figure*}[t]
\begin{center}
\includegraphics[width=.98\textwidth]{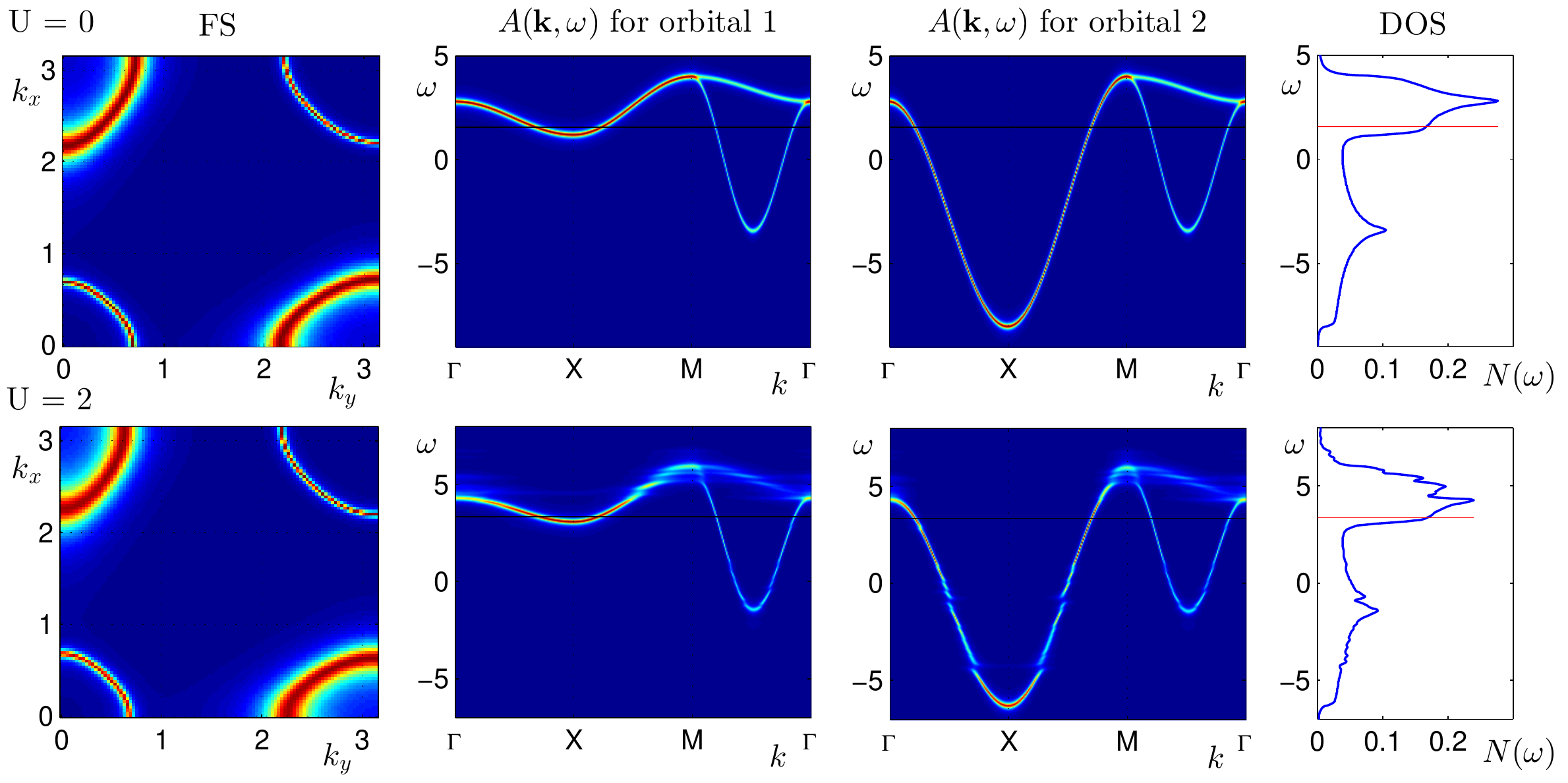}
\caption{FS (on the left), spectral intensity for each orbital $A(\k,\omega)$ (in the middle), and DOS (on the right) for the two-orbital model with zero and finite $U$ are in the top and bottom row, respectively.
\label{fig:specint}}
\end{center}
\end{figure*}

Here we present results of the numerical calculations for Hubbard repulsion $U \leq 2$~eV and Hund's exchange $J = U/4$. Results for larger $U$
will be published elsewhere. Other parameters of calculations are the following: grid in momentum space ($k_x$, $k_y$) is $100 \times 100$, frequency step is $h = 0.05$, and artificial broadening of spectral functions is $\delta = 0.1$. Plots of FS, spectral intensity for each orbital $A_{l l}(\k,\omega)$, and density of states (DOS) for two values of $U$ are shown in Fig.~\ref{fig:specint}. Spectral functions become a little broader away from the Fermi level with the increasing interaction. Apart from that changes to the band structure and the FS are small. This is similar to the results of the variational cluster approximation (VCA) for the two-orbital model with the small $U$~\cite{Daghofer2008,Yu2009}.

\begin{figure}[t]
\begin{center}
\includegraphics[width=.5\textwidth]{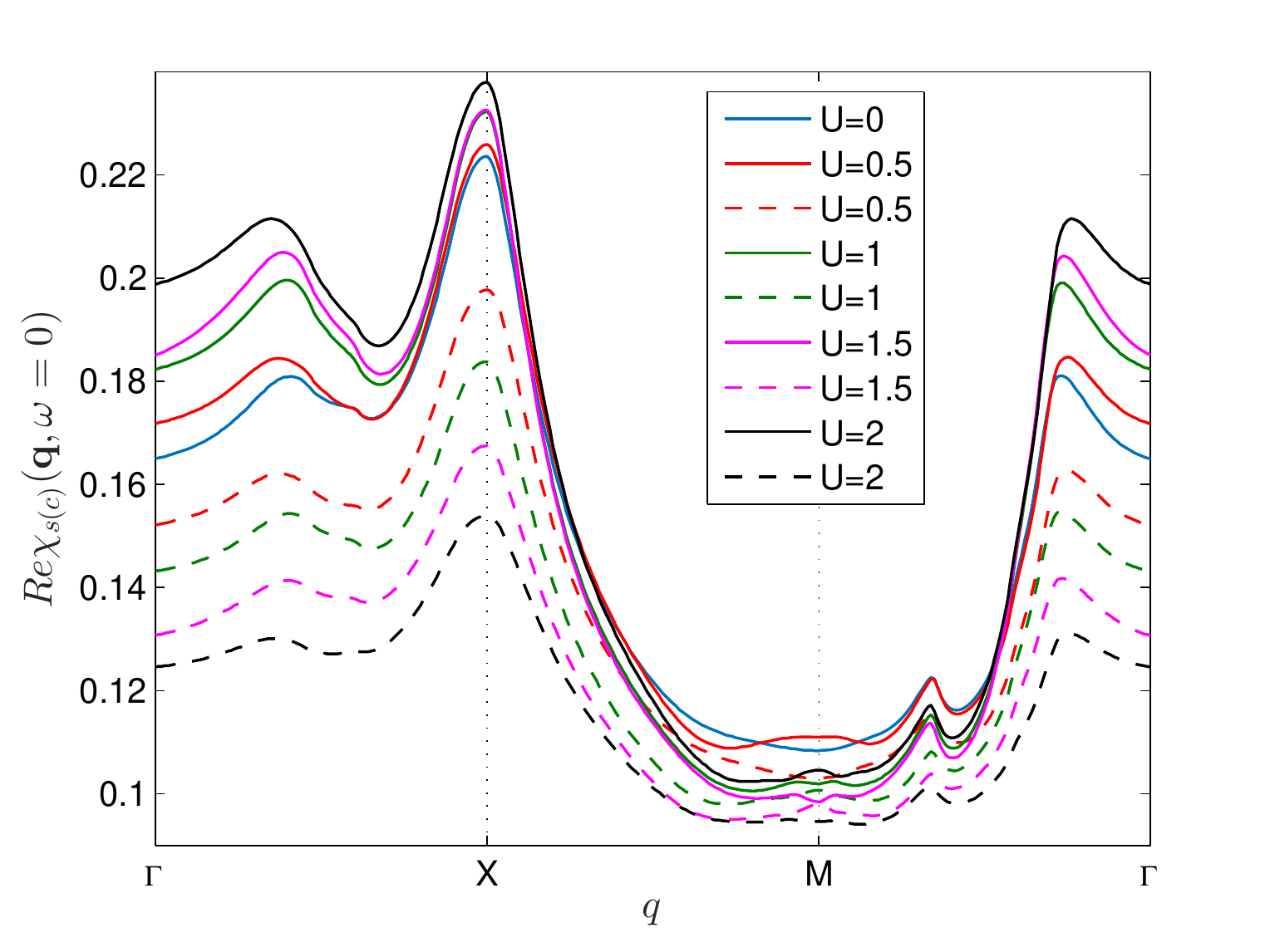}
\caption{Momentum dependence of the real part of physical spin and charge susceptibilities at zero frequency for different values of the Hubbard repulsion $U$. Solid line denotes the spin susceptibility $\chi_{s}(\q,\omega)$, and dashed line denotes the charge susceptibility $\chi_{c}(\q,\omega)$.
\label{fig:rechi}}
\end{center}
\end{figure}

Calculated spin and charge susceptibilities are shown in Fig.~\ref{fig:rechi}. With increasing $U$ the overall magnitude of $\mathrm{Re}\chi_{s}(\q,\omega)$ also increases, while the overall magnitude of $\mathrm{Re}\chi_{c}(\q,\omega)$ decreases. Such behaviour is easily understandable if one recalls that RPA expressions for susceptibilities in a single-band case are $\chi_s = \chi_0 / \left(1 - U \chi_0 \right)$ and $\chi_c = \chi_0 / \left(1 + U \chi_0 \right)$. In the multiband case, these equations should be treated as matrix expressions but obviously the overall trend with changing $U$ will be similar.

\section{Conclusions}

We developed an approximation for calculating spin and charge susceptibilities of a multiband system with the on-site two-body interaction. It combines cluster perturbation theory for the self-energy corrections and RPA for the vertex renormalizations. Calculations for the small values of Hubbard repulsion $U \leq 2$~eV revealed negligible changes in the band structure, FS, DOS, and susceptibilities. This essentially confirms that the rigid band approximation and the RPA for the spin and charge susceptibilities are suitable approaches in the case of weak interactions.

\begin{acknowledgements}
We acknowledge partial support by RFBR (grants 12-02-31534, 13-02-01395, and 16-02-00098) and Government Support of the Leading Scientific Schools of the Russian Federation (NSh-7559.2016.2).
\end{acknowledgements}

\end{document}